\documentclass[12pt]{article}

\usepackage{authblk}
\usepackage{indentfirst}
\usepackage{lipsum}
\usepackage{hhline}
\usepackage{multirow}
\usepackage{tabularx}
\usepackage{etoolbox}
\AtBeginEnvironment{tabular}{\doublespacing}
\usepackage{setspace,caption}
\AtBeginCaption{\doublespacing}
\usepackage{amssymb,amsmath,latexsym}
\usepackage{blindtext}
\usepackage{tabularx,ragged2e,booktabs,caption,subcaption}
\usepackage{graphicx}
\usepackage{booktabs}
\usepackage{lipsum}
\usepackage[justification=justified]{caption}
\usepackage{comment}
\usepackage[width=1.1\textwidth]{caption}
\usepackage[width=0.85\textwidth]{subcaption}
\usepackage[font=footnotesize]{caption}
\usepackage[font=footnotesize]{subcaption}
\usepackage{floatrow}
\usepackage{float}
\usepackage{blindtext}
\usepackage{array}
\usepackage[margin=1in]{geometry}
\usepackage{setspace}
\usepackage{algorithm,algorithmic,xfrac}
\usepackage{lscape}
\usepackage{parskip}
\begin{document}

\title{
       Adapting censored regression methods to adjust for the limit of detection in the calibration of diagnostic rules for clinical mass spectrometry proteomic data}
\author[1*]{Alexia Kakourou}
\author[2]{Werner Vach}
\author[1]{Bart Mertens}
\affil[1]{Department of Medical Statistics and Bioinformatics, Leiden University Medical Center, 2300 RC Leiden, The Netherlands
\\
\texttt{a.a.kakourou@lumc.nl}
\\
\texttt{B.Mertens@lumc.nl}
}

\affil[2]{Center for Medical Biometry and Medical Informatics, University of Freiburg, \\D-79104 Freiburg, Germany
\\
\texttt{wv@imbi.uni-freiburg.de}}
\date{}
\maketitle
\thispagestyle{empty}

\newpage
\pagenumbering{arabic}
\abstract{Despite the recent advances in mass spectrometry (MS), summarizing and analyzing high-throughput mass-spectrometry data remains a challenging task. This is, on the one hand, due to the complexity of the spectral signal which is measured, and on the other, due to the limit of detection (LOD). The LOD is related to the limitation of instruments in measuring markers at a relatively low level. As a consequence, the outcome data set from the quantification step of proteomic analysis often consists of a reduced list of peaks where any peak intensities below the detection limit threshold are reported as missings.
In this work, we propose the use of censored data methodology to handle spectral measurements within the presence of LOD, recognizing that those have been censored due to left-censoring mechanisms on low-abundance proteins. We apply this approach to the particular problem of calibrating prediction rules through prior estimation of the average isotope expression in MALDI-FTICR mass-spectrometry data, collected in the context of a pancreatic cancer case-control study. Our idea is to replace the set of incomplete spectral measurements with the average intensity estimates and use those as new input to a prediction model. We evaluate the proposed methods, with respect to their predictive ability, by comparing their performance with the one achieved using the complete information as well as alternative/competitive methods to deal with the LOD.
\\
\mbox{}
}

\newpage

\section{Introduction}

One of the primary objectives in mass-spectrometry (MS) clinical proteomics is to detect and quantify the proteomic expression present in biological samples. However, the quantification step in the analysis of mass-spectrometry data can be hampered by the fact that measurements may be subject to lower detection limits due to censoring mechanisms on low-abundance proteins/peptides \cite{karp1,karp2}. This issue is known as limit of detection (LOD) and it occurs due the limited ability of instruments to measure low-concentration spectral features. As a result, the acquired proteomic data set, in many applications, consists of a reduced list of peaks in which any peak intensities below the detection limit threshold are reported as missing values.

Several approaches have been proposed in the literature to deal with data subject to lower or upper detection limits, particularly in ecological and environmental research \cite{dennis1,dennis2,rubin,reed,chen}. The topic of handling proteomic data within the presence of LOD has recently emerged in the field of mass-spectrometry clinical proteomics. 
Dong et al. \cite{dong} addressed the problem of assessing bias in the estimation of distribution parameters of proteomic biomarkers whose measurements were subject to the LOD. In their paper they considered a protein pathway data set and proposed methods to combine proteomic markers, adjusting for the LOD, in order to distinguish cancer patients from non-cancer patients. They showed that the ROC curve parameter estimates generated from the proposed methods are much closer to the truth as compared to simply combining proteomic markers ignoring the LOD. Tekwe et al. \cite{tewke} on the other hand acknowledged the LOD issue in (MS) proteomic data as a problem of censored data analysis and proposed the use of survival methodology, and in particular accelerated failure time models (AFT), to carry out differential analysis of proteins. They proved that AFT models have higher ability to detect differentially expressed proteins than standard testing procedures, with the discrepancy widening with increasing missingness in the proportions.

In this paper we further adapt the use of censored data methodology to handle spectral measurements within the presence of LOD. We implement this approach to the particular problem of estimating the proteomic expression within isotopic clusters with the ultimate goal of using the newly derived estimates as new predictor variables for the calibration of diagnostic rules. In particular, we adapt censored normal regression methods to estimate the average intensity within an isotope cluster for each individual, based on partially observed MALDI-FTICR mass-spectrometry data, collected in the context of a pancreatic cancer case-control study. The estimates of average, adjusted for LOD, intensity are later used as input to a prediction model. While censored regression models are widely used in the survival analysis field, the specific case of censored normal regression, as considered in this paper, is often used in econometrics to handle skewed data and it is referred to as Tobit regression \cite{tobin}. We further combine censored regression with borrowing of information, through the addition of an individual-specific random effect formulation, to correct for both lack of information and measurement uncertainty. In addition we present an extension which allows for selection of a subset of features based on the parameter estimates of the censored model. We evaluate the proposed methods, with respect to their predictive ability, by comparing their performance with the one achieved using the complete information as well as alternative\hskip0pt/\hskip0ptcompetitive methods to deal with the LOD.


The remainder of the paper is organised as follows: In section 2 we give a brief overview of the data and the data structure. In section 3 we present different frameworks of using random effects censored regression to estimate the average isotope expression in the individual spectra for prediction purposes. Section 4 contains results and relative performance of the new/proposed approaches with methods of ad-hoc nature. Additionally, we show how the use of individual-specific random effects in the censored regression model may allow for the selection of fairly sparse models while maintaining good predictive performance. We finish with a discussion in Section 5.

\section{Data}
\subsection{Data description}
We consider data from a case-control study, the design of which is described in detail in Nicolardi et al. \cite{simone}. For the experiment, serum samples were collected from 88 patients with pancreatic cancer and 185 healthy volunteers. The samples from the included individuals were stored and processed according to a standardized protocol. The study design defined a calibration set and a separate validation set. The validation samples were collected in a later time period. For the calibration set, serum samples were obtained from 49 pancreatic cancer patients and 110 healthy controls (age- and gender- matched) while for the validation set samples were obtained from 39 pancreatic cancer patients and 75 healthy (age- and gender- matched) controls. The available calibration and validation samples were distributed over three distinct MALDI-target plates and were mass-analysed by a MALDI-FTICR MS system resulting in a single spectrum per sample covering the mass/charge range from 1013 to 3700 Da.

In Figure 1 we plot the mass spectrum for a case sample. A mass spectrum consists of peaks with a certain intensity distributed over a \textit{m/z}-axis generated from the detection of ionized molecules. In ultrahigh-resolution mass spectrometry, each species (such as peptide) is detected as a cluster of peaks. These peaks represent ions of the same elemental composition but different isotopic composition due to the presence of additional neutrons and form the so-called isotopic cluster. Superimposed in Figure 1 is shown an isotopic cluster at position \textit{m/z} $2021.2$.

\subsection{Data structure and limit of detection (LOD)}

We apply to the complete raw spectra a peak detection algorithm \cite{alexia} using a fixed LOD threshold which reflects the background noise level in the individual spectra. This results in a reduced list of peaks which is a data format typically encountered in proteomics research when full spectral measurements are reduced to a discrete set. In case a peak is observable/detectable in a patient, the approach calculates the area under the intensity curve to obtain an intensity value for that peak and that patient. In case a peak is unobservable/undetectable in a patient, we regard the intensity value for that peak and that patient as left-censored due to the LOD.
Our objective is to investigate whether, starting from the reduced data, we can develop methods to recuperate information on the average individual expression within clusters which will allow us to calibrate diagnostic rules of comparable performance as if we had the complete information.


The resulting list of peaks for the pancreatic cancer data contained $8080$ identified isotopic peaks dived into $2717$ identified isotopic clusters. Only a small number of these peaks was observable/detectable in all samples. The output data set contained hence a large proportion of censored intensity values (85 \%).
The structure of the observed data, given the incomplete response due to the LOD, is given by $(G_i,\mathbf{Y}_i,\mathbf{\Delta}_i)$, $i=1,...,n$, where $G$ is the vector containing the group outcome, $\mathbf{Y}$ is a matrix of dimension $n \times p$, containing the quantified peak intensity values and $\mathbf{\Delta}$ is a matrix of dimension $n \times p$, representing the censoring indicators which take the value $\delta_{ij}=1$, if peak $j$ is observable, and $\delta_{ij}=0$, if peak $j$ is unobservable in the $i$th individual. In the latter case, we set the value of $y_{ij}$ to the LOD value.

\section{Methods}
%
%
In this section we consider methods to construct, for a given cluster, estimates $\hat{\bar{y}}_{i}$ of the average intensity based on the log-transformed peak intensities $y_{ij}$, with $i=1,...,n$ denoting the patients and $j=1,...,k$ denoting the peaks of the cluster. These estimates will later be used as new input variables for the construction of a diagnostic rule. Our main approach will be based on a simple model for the intensities in a single patient, which takes into account the fact that we can observe a common pattern in the intensities across patients. More specifically, we postulate a regression model for the intensities of a patient in a cluster, using the empirical pattern of mean intensities $\bar{y}_j := \frac{1}{n} \sum_{\substack{i}} y_{ij}$ across patients as covariate. To obtain the estimates of the average intensity for each patient and each cluster, we use censored regression methodology.

In addition to the censored regression models, we consider some well-known strategies which can deal with unobservable intensity values at the peak level. The first and simplest alternative strategy we consider is complete case analysis, ignoring thus all censored peak intensities.
A simple alternative approach which allows us to use additional information on the unobservable peaks is to reduce all intensity values to the binary information above/below the LOD.
Finally, we consider substituting the unobservable peak intensities with the LOD value in order to avoid the loss of information in the observable peaks. For all these alternative methods, we obtain an estimate of the average intensity within cluster by averaging the (available) values. 
For the complete case analysis, if an entire cluster is unobservable in a sample, 
we impute the average over the estimates from all patients with at least one observable peak.

\subsection{Censored regression}

We consider for each patient and each cluster a simple regression model for the true log-transformed intensities, of the type
\[ \tilde{y}_{ij} = \alpha_i + \beta_i \bar{y}_j + \varepsilon_{ij} \]
\[\varepsilon_{ij}\sim N(0,\sigma_i^2) \]

The model parameters $\alpha_i$ and $\beta_i$ in the above expression capture the intensity variation of a particular individual. Here, $\alpha_i$ reflects the systematic differences in average expression across patients while $\beta_i$ represents the (multiplicative) effect of the average isotope pattern $\bar{y}_j$, which is expected to be a rather good predictor of the observed pattern in each individual. The likelihood function based on the partially censored observations is given by
\[ L(\theta_i)=\prod\limits_{j=1}^{k} f(y_{ij},\theta_i)^{\delta_{ij}} F(y_{ij},\theta_i)^{1-\delta_{ij}} \]
where $\theta_i$ is the vector of model parameters, $f(y_{ij},\theta_i)$ is the probability density function of the normal distribution and $F(y_{ij},\theta_i)$ is the cumulative density function. The contribution of the observed peak intensities to the likelihood is given by $f(y_{ij},\theta_i)$ while the contribution of the left-censored peak intensities is given by $F(y_{ij},\theta_i)=Pr(y_{ij}\le t)$ where $t$ denotes the minimum detectable threshold. The estimates of the regression parameters are obtained by maximizing the log-likelihood function
\[ l(\theta_i)=\sum\limits_{j=1}^k \delta_{ij}\log f(y_{ij},\theta_i) +(1-\delta_{ij})\log F(y_{ij},\theta_i)\]

We summarize the entire isotopic expression within each cluster and for each individual as a function of the estimates $\hat{\alpha}_i$ and $\hat{\beta}_i$, given by
\[ \hat{\bar{y}}_i=\frac{1}{k} \sum\limits_{j=1}^k \big(\hat{\alpha}_i+\hat{\beta}_i\bar{y}_j\big)=\hat{\alpha}_i+\hat{\beta}_i\bar{\bar{y}} \]
Finally, we use the set of the derived estimates $\{ \hat{\bar{y}}_i, \enspace i=1,...,n \}$ as new predictor variables for the construction of the discriminating rule.

\subsection{Random effects censored regression}
\indent Depending on the extent of left-censoring, information in a cluster for a specific patient may either be insufficient for estimating the model parameters or include great uncertainty resulting in unreliable parameter estimates. We account for lack of information and measurement uncertainty by combining censored regression with shrinkage estimation of the intensity levels. The key idea is to adjust the estimates of the less reliable individual expressions in a cluster by pooling information across all available patients. In analogy to repeated measures data analysis, we treat the peak intensities within each cluster for each patient as repeated observations and we fit for each cluster a joint model across patients including individual-specific random effects.

We restrict our investigation to a simple univariate random effects model, given by
\[ \tilde{y}_{ij} =  a_i + \alpha  + \beta \bar{y}_j  + \varepsilon_{ij} \]
\[\varepsilon_{ij}\sim N(0,\sigma^2)\]
\[ a_i \sim N(0,\tau^2)\]
In the above model specification we choose a fixed effect representation for $\beta$, primarily due to computational constraints when fitting a bivariate random effects model, as the fitting process requires numerical integration based on summing up over a number of fixed points which grows exponentially with the number of dimensions. 
In this way, $\alpha$ represents the mean intercept across all patients while $a_i$ represents the individual deviation from the mean. The variation of the individual intercepts around the mean is assumed to be normally distributed.
\\
\indent The likelihood function for the parameter vector $\theta=\big(\alpha,\beta,\tau^2,\sigma^2\big)$, adjusted for the LOD, is given by
\[ L(\theta;\textbf{y}_i)= \prod_{i=1}^n  \Bigg(\int^{+\infty}_{-\infty} \prod_{j=1}^k \Big( f(y_{ij}|a_i)^{\delta_{ij}}F(y_{ij}|a_i)^{1-\delta_{ij}}\Big)f(a_i) \,da_i   \Bigg) \]
By conditioning on the random effect $a_i$, the above expression for the likelihood function specifies that a detectable peak intensity $y_{ij}$ contributes $f(y_{ij}|a_i)$ whereas an undetectable peak intensity contributes $F(y_{ij}|a_i)$, i.e. a Bernoulli probability that $y_{ij}$ is below the minimum detectable threshold $t$. When undetectable peak intensities are to be accounted for, the Bayes estimate of the random effect $a_i$ can be computed by substituting the ML estimates of $\theta=\big(\alpha,\beta,\tau^2\sigma^2\big)$ into the analytic expression for the posterior mean given the observed data, given by
\[ E(a_i|\textbf{y}_i,\textbf{$\theta$})= \frac{1}{f^{\ast}(\textbf{y}_i;\theta)}\int^{+\infty}_{-\infty} a_i \ f^{\ast}(\textbf{y}_i|a_i)f(a_i) \,da_i\]
where,
\[ f^{\ast}(\textbf{y}_i,\theta)= \int^{+\infty}_{-\infty}   f^{\ast}(\textbf{y}_i,a_i) da_i =
                                  \int^{+\infty}_{-\infty}   f^{\ast}(\textbf{y}_i |a_i)f(a_i) \,da_i \]
The asterisk denotes the fact that the intensity vector $\textbf{y}_i$ may include one or more undetectable values. The above expression is equivalent to
\[ \hat{a}_i=E(a_i|\textbf{y}_i,\textbf{$\theta$})=\frac{\tau^2}{\tau^2+\sfrac{\sigma^2}{k}}\Big(E\big(\bar{\tilde{y}}_i | \textbf{y}_i,\boldsymbol\delta_i,\hat{\theta}\big)-\big(\hat{\alpha}+\hat{\beta}\textbf{$\bar{y}$}\big)\Big) \]
as pointed out by Hughes \cite{hug}, who proposed prediction of random effects in conjunction with an EM approach to the LOD problem. Finally, the estimated intensity for the $i$th individual can be written as
\begin{equation*}
\begin{split}
\hat{y}_i & =  \hat{\alpha} + \hat{\beta} \bar{y} + \hat{a}_i \\
 & =  \hat{\alpha} + \hat{\beta} \bar{y} + \frac{\hat{\tau}^2}{\hat{\tau}^2+\sfrac{\hat{\sigma}^2}{k}}\Big(E\big(\bar{\tilde{y}}_i | \textbf{y}_i,\boldsymbol\delta_i,\hat{\theta}\big)-\big(\hat{\alpha}+\hat{\beta}\textbf{$\bar{y}$}\big)\Big) \\
 & = \bigg(1-\frac{\hat{\tau}^2}{\hat{\tau}^2+\sfrac{\hat{\sigma}^2}{k}} \bigg)\big(\hat{\alpha} + \hat{\beta} \bar{y}\big)+ \bigg(\frac{\hat{\tau}^2}{\hat{\tau}^2+\sfrac{\hat{\sigma}^2}{k}}\bigg) E\big(\bar{\tilde{y}}_i | \textbf{y}_i,\boldsymbol\delta_i,\hat{\theta}\big)
\end{split}
\end{equation*}
Viewed this way, the estimate of $\tau^2$ plays the role of a shrinkage parameter which, depending on the available information in a cluster, pulls the estimate of a particular individual to a greater or a smaller extent towards the population mean. Note that $\hat{y}_i$ is defined also in the case of completely unobservable clusters, hence patients for which all intensity values are censored in a cluster can also be included in the analysis.

\subsection{Random censored regression applications}
There are several possibilities of using the censored regression model with random effects to estimate the individual isotopic expression (while accounting for the LOD) with the ultimate goal being to use the newly derived estimates as new predictor variables for the calibration of the diagnostic rule. In the following we demonstrate three different variants of using the random censored model, for prediction purposes.

\subsubsection{Random censored regression as a preprocessing tool}
A simple and straightforward way to summarize the incomplete predictive information in the pancreatic data, while accounting for the LOD, is to apply the random effect censored regression approach across all available data i.e. data from both calibration and validation sets. This can be considered as a means of preprocessing the data, without using information on the class outcome, prior to building the diagnostic rule.

The Bayes estimate of the random intercept in this case is given by $\hat{a}_i(\hat{\theta}_{\textit{all}})=E(a_i|\textbf{y}_i,\textbf{$\theta$}_{all})$ where $\hat{\theta}_{\textit{all}}=(\hat{\alpha}_{\textit{all}},\hat{\beta}_{\textit{all}},\hat{\tau}_{\textit{all}}^2,\hat{\sigma}_{\textit{all}}^2)$ and \textit{all} denotes the fact that the estimates of the parameter vector were derived based on all the available observations. Correspondingly, the expected intensity for patient $i$ and peak $j$ is given as
\[\hat{y}_{ij}(\hat{\theta}_{\textit{all}})=\hat{a}_{i}(\hat{\theta}_{\textit{all}}) + \hat{\alpha}_{\textit{all}}+ \hat{\beta}_{\textit{all}} \ \bar{y}_j\]
while the average expected intensity within cluster for patient $i$ is derived by
\[\hat{\bar{y}}_i(\hat{\theta}_{\textit{all}})=\hat{a}_{i}(\hat{\theta}_{\textit{all}}) + \hat{\alpha}_{\textit{all}}+ \hat{\beta}_{\textit{all}} \ \bar{\bar{y}}\]

\subsubsection{Random censored regression as a prediction tool}
A more formal approach, more in tune with predictive calibration and subsequent validation, is to embed the above estimation procedure within the ordinary prediction framework. This suggests using the calibration data for both parameter estimation of the random censored model and construction of the prediction model and subsequently applying the resulting rules to the set-aside validation set.

In that case, the Bayes estimate of the random intercept is given by $\hat{a}_i(\hat{\theta}_{\textit{cal}})=E(a_i|\textbf{y}_i,\textbf{$\theta$}_{cal})$ with $\hat{\theta}_{\textit{cal}}=(\hat{\alpha}_{\textit{cal}},\hat{\beta}_{\textit{cal}},\hat{\tau}_{\textit{cal}}^2,\hat{\sigma}_{\textit{cal}}^2)$ where \textit{cal} denotes the fact that the parameter estimates were based solely on the calibration samples. In other words, both calibration and validation data are shrunken according to the estimates derived based on the calibration set alone. The expected intensity of each patient $i$, for the calibration and the validation sets, is given by
\[ \hat{y}_{{ij}_{\textit{cal}}}(\hat{\theta}_{\textit{cal}})=\hat{a}_{i_{\textit{cal}}}(\hat{\theta}_{\textit{cal}}) + \hat{\alpha}_{\textit{cal}} + \hat{\beta}_{\textit{cal}} \ \bar{y}_j \quad
\text{and} \quad
\hat{y}_{{ij}_{\textit{val}}}(\hat{\theta}_{\textit{cal}})=\hat{a}_{i_{\textit{val}}}(\hat{\theta}_{\textit{cal}}) + \hat{\alpha}_{\textit{cal}} + \hat{\beta}_{\textit{cal}} \ \bar{y}_j\]
respectively while the corresponding cluster summary for each set is given by
\[ \hat{\bar{y}}_{{i}_{\textit{cal}}}(\hat{\theta}_{\textit{cal}})=\hat{a}_{i_{\textit{cal}}}(\hat{\theta}_{\textit{cal}}) + \hat{\alpha}_{\textit{cal}} + \hat{\beta}_{\textit{cal}} \ \bar{\bar{y}} \quad
\text{and} \quad
\hat{\bar{y}}_{{i}_{\textit{val}}}(\hat{\theta}_{\textit{cal}})=\hat{a}_{i_{\textit{val}}}(\hat{\theta}_{\textit{cal}}) + \hat{\alpha}_{\textit{cal}} + \hat{\beta}_{\textit{cal}} \ \bar{\bar{y}}\]

\subsubsection{Random censored regression re-estimation}

Estimating the average intensity in the validation data, either in conjunction with the calibration data, as in the case of the random censored regression model as a preprocessing tool, or according to the censored regression estimates based on the calibration data, as in the case of the random censored regression model as a prediction tool, implies that one assumes that both calibration and validation data stem from the same population.

If the above assumption does not hold it may be necessary and/or beneficial to estimate the censored model parameters separately for the calibration and validation data since difference in population could result in potentially different values of $\alpha$, $\beta$, $\tau^2$ or $\sigma^2$ between the two sets. This could be particularly true in the case of external validation where validation samples may represent a different population than calibration samples.

On the assumption that the two populations are different, we fit the random effect censored regression model separately to the calibration and the validation data. In this case, the random effect estimate for the calibration samples is given by
$\hat{a}_{i_{\textit{cal}}}(\hat{\theta}_{\textit{cal}})=E(a_{i_{\textit{cal}}}|y_{i_{\textit{cal}}},,\textbf{$\theta$}_{cal})$, where $\theta_{\textit{cal}}=\big(\alpha_{\textit{cal}},\beta_{\textit{cal}},\tau_{\textit{cal}}^2,\sigma_{\textit{cal}}^2)$, while the random effect estimate for the validation samples is given by $\hat{a}_{i_{\textit{val}}}(\hat{\theta}_{\textit{val}})=E(a_{i_{\textit{val}}}|y_{i_{\textit{val}}},,\textbf{$\theta$}_{val})$, where $\theta_{\textit{val}}=\big(\alpha_{\textit{val}},\beta_{\textit{val}},\tau_{\textit{val}}^2,\sigma_{\textit{val}}^2)$. The resulting intensity estimates for the calibration and validation sets are derived by
\[ \hat{y}_{{ij}_{\textit{cal}}}(\hat{\theta}_{\textit{cal}})=\hat{a}_{i_{\textit{cal}}}(\hat{\theta}_{\textit{cal}}) + \hat{\alpha}_{\textit{cal}} + \hat{\beta}_{\textit{cal}} \ \bar{y}_j \quad
\text{and} \quad
\hat{y}_{{ij}_{\textit{val}}}(\hat{\theta}_{\textit{val}})=\hat{a}_{i_{\textit{val}}}(\hat{\theta}_{\textit{val}}) + \hat{\alpha}_{\textit{val}} + \hat{\beta}_{\textit{val}} \ \bar{y}_j\]
respectively while their corresponding cluster summaries are defined as
\[ \hat{\bar{y}}_{{i}_{\textit{cal}}}(\hat{\theta}_{\textit{cal}})=\hat{a}_{i_{\textit{cal}}}(\hat{\theta}_{\textit{cal}}) + \hat{\alpha}_{\textit{cal}} + \hat{\beta}_{\textit{cal}} \ \bar{\bar{y}} \quad
\text{and} \quad
\hat{\bar{y}}_{{i}_{\textit{val}}}(\hat{\theta}_{\textit{val}})=\hat{a}_{i_{\textit{val}}}(\hat{\theta}_{\textit{val}}) + \hat{\alpha}_{\textit{val}} + \hat{\beta}_{\textit{val}} \ \bar{\bar{y}}\]

\section{Application and analysis}

\subsection{Model choice}
We assess the performance of the proposed methods by fitting a prediction model to the set of the derived cluster summaries and by evaluating the predictive performance of each such fit. Summarizing the isotopic expression per cluster results in a total of 2717 cluster summaries, reducing thus the dimensionality of the original predictor data to a lower dimensional space. In this lower-dimensional space, the number of predictor variables still exceeds the number of observations, therefore we choose to use ridge logistic regression \cite{ridge} to calibrate the diagnostic rule. This method is proved to be very effective in high-dimensional settings, where the number of covariates exceeds the number of observations and/or there are high correlations between them. Ridge regression deals with overfitting and collinearity by maximizing the log-likelihood function with a penalty on the regression coefficients.

\subsection{Model fitting and performance measures}
To fit the random intercept censored regression model we use the NLMIXED procedure available in SAS which is written to fit non-linear mixed models. The computation of the integral over the random effect is performed by an adaptive Gaussian quadrature method with 100 integration points.
Since in our data analysis we consider integrated intensities, the original LOD value used in our peak detection algorithm is no longer adequate, as it only points out/indicates that the maximal intensity in an interval around the peak is below that value. Taking into account that the width and shape of a particular peak is rather constant across patients, we decided to use as peak specific LOD value, the minimal observed integrated intensity among all patients with an uncensored measurement.

To evaluate the proposed methods with respect to their predictive performance, we first apply a type of internal validation in which we use random splitting to redefine the calibration and validation sets. This allows us to assess consistency of performance estimates and obtain more robust results. The new calibration-validation structure is defined in such a way that it respects the case/control ratio of the original study design \cite{alex}. The whole procedure is repeated 10 times and classification results across the different repetitions are finally averaged to obtain more stable estimates.

To assess the predictive performance of the diagnostic rules based on the cluster summaries derived using each random censored regression approach we perform leave-one-out cross validation on the re-defined calibration set to select the optimal ridge penalty and we then apply the resulting discriminating rule to the re-defined validation set \cite{stone,bart}. For each model we calculate the error-rate and the area under the ROC curve (AUC). To evaluate the accuracy of each fit, we also calculate the Brier score, given by
\begin{eqnarray*}%
Brier\;score &=& \frac{1}{n_{val}}\sum\limits_{i=1}^{n_{val}} (\hat{p}_i-g_i)^2
\end{eqnarray*}
and the deviance, defined as
\begin{eqnarray*}
Deviance &=& -2\sum\limits_{i=1}^{n_{val}} g_i\log{\hat{p}_i}+(1-g_i)(\log{(1-\hat{p}_i})
\nonumber\\
&=&-2\sum\limits_{i=1}^{n_{val}} \log(1-|\hat{p}_i-g_i|)
\end{eqnarray*}
where $\hat{p}_i$ is the estimated probability of being a case for the $i^{th}$ validated individual, $g_i$ is the known class outcome of that individual and $n_{val}$ is the total validation sample size. To compute the error-rates we use a threshold of $0.5$ and we assign an observation as a diseased case if the predicted class probability $\hat{p}_i$ is greater than $0.5$ and as a control otherwise.

\subsection{Results}
Table 1 shows validated performance measures, together with standard errors, of the ridge logistic model fitted to the set of average estimates based on complete case analysis (CCA), binary coding (BC), substituting unobservable peak intensities with the detection limit value (LOD), random censored regression as a preprocessing tool (CR Prep), as a prediction tool (CR Pred) and re-estimated (CR Reest).
The last column of Table 1 contains performance measures based on substituting the unobservable peak intensities with the area under the intensity curve in a systematic interval around the peak position, with length corresponding to the typical peak width in a specific \textit{m/z} range, estimated from the raw data. This approach can be considered equivalent to having the complete information on the peak intensities (the ``truth'') and it is feasible in our specific situation since we have access to the complete spectra and not just a peak list as it is often the case. Therefore, assessment of relative performance may be carried out with respect to this approach (TR).

Performance measures, based on BC suggest that the present/absent patterns of the proteomic expression are highly informative with regards to the class outcome.
Incorporating additional information on the relative intensity, while accounting for the LOD, seems to be recovering information on top of the present/absent information. Specifically, results based on estimating the average intensity using CR Prep or CR Pred indicate that using censored regression strategies combined with pooling of information provide a nice solution to the LOD problem, both from a statistical and practical point of view. Performance measures based on CR Reest show no improvement over CR Prep or CR Pred when re-estimating the random censored model estimates in the validation set, in the case of internal validation. Nevertheless, this outcome is anticipated as CR Reest is expected to be optimal only in case we have evidence or prior knowledge about the calibration and validation populations being different.

Results based on CCA illustrate that ignoring the unobservable peak intensities when estimating the average intensity on which the classification rule is based results in poor classification results as compared to results based on estimating the values of the unobservable peaks using an approach designed for censored data.
Interestingly, we observe that substituting the unobservable peak intensities with the LOD value 
results in comparable performance to the one achieved using CR Prep, CR Pred,CR Reest or TR. However, this is not utterly surprising as the LOD value for the pancreatic cancer data is a rather good estimate of the true (unobservable) intensity value.

Next, we apply each method to the original data. This can be regarded as a type of external validation during which we build the diagnostic rule based on the set of average intensity estimates on the calibration set, as defined in the original study, and we evaluate the resulting rule on the set of cluster summaries on the separate validation set. Validated classification results for all methods are shown in Table 2. With one exception, we observe comparable ranking of the methods with the one based on internal validation. Improvement in predictive performance of the proposed censored regression methods as compared to CCA (as well as all other methods including TR) is more apparent in this case, as indicated by both the error-rate and the AUC.
In particular, CR Reest outperforms now all methods (including CR Prep, CR Pred and TR) in all performance measures. This outcome provides some confirmation on the value of the re-estimation approach when the two populations are known to be different, as in the case of our external validation. Investigations which would allow us to gain more insight into the possible situations under which the re-estimation approach is expected to outperform the alternative strategies is left as an interesting line of future research.

Finally, we explore to which degree the achieved classification performance when using random censored regression as a solution to the LOD problem is due to borrowing of information or due to shrinkage of the level estimates.
We address this question by fitting a prediction model with $E\big(\bar{\tilde{y}}_i | \textbf{y}_i,\boldsymbol\delta_i,\hat{\theta}\big)$ as input variable. In case of no censoring, the above expression reduces to the observed average intensity within the cluster. In this way we allow for ``borrowing'' in estimating the average intensity of a cluster without using shrinkage.
Performance measures using the ``unshrunken'' estimates derived based on CR Prep, CR Pred and CR Reest for the internal and external validations are shown in Table 3. 
When comparing these results to the ones from Tables 1 and 2, we observe that, in the case of internal validation, using censored regression methods without shrinkage often results in similar classification performance as using censored regression with shrinkage. This suggests that the use of censored regression provides a solution to the LOD problem also when it is not combined with shrinkage. In the case of external validation, we observe larger discrepancies favouring the use of shrinkage. This outcome suggests that there may be situations where using shrinkage has a value in its own right.


\subsection{Variable selection}
A nice property of combining censored regression with borrowing of information is that it may allow for some type of variable selection, based on the estimate of the random effect variance. As already discussed in Section 3.2, the variance of the cluster-specific random intercept $\tau_c^2$ acts as a shrinkage parameter. Depending on the amount and reliability of the available information in cluster, the estimates of a specific individual are pulled to a smaller or a greater extent, towards the common population mean. Accordingly, the larger the value of $\tau_c^2$, the higher the spread from patient to patient and hence the more informative that cluster may be. The above consideration can be used as a criterion to eliminate clusters with minimal $\tau_c^2$. 
Already, the random effect variances for $15$ clusters were estimated as $0$ by the CR Prep approach and thus these clusters were automatically ignored by ridge regression.

Selecting only a subset of clusters, while maintaining the good predictive performance, can be of particular interest for potentially measuring solely proteins/peptides at predefined \textit{m/z} locations, reducing thus the cost of measurement and storage for future data and facilitating all subsequent analyses. Moreover, variable selection may allow for the identification of a set of features which are likely to be associated with the disease mechanisms and disease progression and therefore could provide leads to further exploit diagnostic and therapeutic potential.

\subsubsection{Variable selection prior to calibration}
A simple way to perform variable selection is to discard a certain fraction of clusters with minimal $\tau_c^2$. For instance, we may decide to eliminate 50\%, 80\% or 90\% of the total number of clusters with minimal $\tau_c^2$ prior to calibrating the diagnostic rule. In this way, the decision on which clusters to retain or omit depends solely on the magnitude of the random effect variance and not on cross-validated risk looking at the class outcome. The first 4 boxplots of Figure 2 represent error-rate distributions when keeping all clusters, 50\%, 20\% and 10\% of total clusters with minimal $\tau_c^2$, as estimated by CR Prep (upper plot), CR Pred (middle plot) and CR Reest (lower plot) for the 10 re-sampled validation sets. These results suggest that we can omit at least half of the clusters from the analysis without deteriorating the predictive performance. Note that for CR Reest, though we get different regression estimates for the calibration and validation sets, the decision on which clusters to omit (in both calibration and validation sets) is again based solely on the random effect variance estimate of the calibration set.

\subsubsection{Variable selection within calibration}
Since we are in the prediction setting, we may choose the optimal fraction of clusters to retain or omit directly from the predictive perspective, by optimizing the cross-validated loss function with respect to predictive performance in the usual cross-validatory way. We do so by considering the fraction of selected clusters $F$ as a tuning parameter to be optimized. In this case, estimation involves combined optimization of the fraction $F$ and the ridge penalty $\lambda$. To optimize $F$ (in conjunction with $\lambda$) we perform leave-one-out cross-validation on the calibration set for a grid of $20$ $F$ values corresponding to the ventiles of $\boldsymbol{\tau}^2=(\tau_1^2,...,\tau_C^2)$.

Figure 3 shows an example of the cross-validated error, for CR Prep and the first random split, as the fraction of selected clusters $F$ becomes smaller. 
The cross-validated error is minimized at $F=20\%$, resulting in a subset of just 543 clusters/proteins. 
If the curve is flat near the minimum, we choose the smallest fraction that achieves the minimal error, favoring hence sparser models. The error-rate distribution based on optimised $F$ for the 10 random splits is shown in the last boxplot of Figure 2. With one exception, $F$ was consistently estimated as either 30\% or 20\% across the random splits, with hardly any loss of predictive accuracy as it can be seen from Figure 2.

\section{Discussion}

\indent In this paper we proposed to adapt censored regression methods to estimate the average individual expression within isotopic clusters, prior to building prediction rules, as a way to deal with the limit of detection. We evaluated the proposed methods, with respect to predictive performance, by replacing the incomplete spectral measurements with the newly derived estimates of individual expression, accounted for the LOD, and using those as new predictor variables for the construction of diagnostic rules. We combined censored regression with borrowing of information across data to account for potential lack of information and measurement uncertainty. Results from both internal and external validations indicated that using the estimates from the proposed methods as new input variables results in comparable predictive accuracy to the one achieved using the complete intensity information. Ignoring the unobservable peak intensities, as an alternative to deal with the LOD, resulted in poor predictions as compared to the proposed methods while substituting the unobservable peak intensities with the LOD value exhibited similar classification performance as the proposed methods.
\\
\indent We demonstrated different variants of using censored regression, in combination with borrowing of information, for prediction purposes. Random censored regression as a preprocessing tool is straightforward in application since, at a first instance, it only requires fitting the random censored model across all available data to obtain the adjusted for LOD estimates of individual expression. However, since the derived estimates of a particular individual depend now on the expression from all other individuals due to the explicit borrowing of information, information from the validation samples enters the rule derived based on the calibration samples. Since our objective is prediction, we may choose to avoid this by using only the calibration samples to fit the censored regression model and use the derived estimates to adjust for the LOD in the validation set. This approach respects the formal prediction framework.
\\
\indent Another aspect related to the above comparative discussion between censored regression as a preprocessing or prediction tool is the potential need of re-estimating the random censored model parameters in the validation set when the samples represent a different population than the calibration samples, as in the case of external validation. In classification problems the issue of population difference may not be as crucial, since we model the conditional distribution of the class outcome. 
However, in data preprocessing, where we model each single univariate covariate, population difference could be of consequence as in that case fitting the random censored model separately to the calibration and validation data could potentially lead to distinct parameter estimates for the two different data sets.
Results based on our external validation provided rather clear evidence that the re-estimation approach is optimal in this case. Nevertheless, further investigations are required to find out in which situations and to which degree one can benefit from choosing this specific variant of using censored regression methods with random effects to account for the LOD.
\\
\indent We restricted our discussion to the simple case of the univariate random effects model with random intercept only. We chose to use the univariate random effects model due to its relative ease in computation as opposed to the bivariate case with both random intercept and slope, as fitting these models requires numerical integration based on summing up over a number of fixed grid points which grows exponentially with the number of dimensions.
In fact, it might be of interest to consider the bivariate random effects model since the degree to which the average pattern is predictive of the observed pattern may vary from patient to patient. However, results based on using the estimates from a bivariate random censored model (as a preprocessing tool) as predictors were identical to those based on using the estimates from the univariate random censored model (as a preprocessing tool), suggesting that incorporating this additional information is insufficient to allow for improved predictions. Moreover, for a large number of clusters, the estimate of the random slope variance was close to zero, justifying thus the choice of keeping the slope fixed. Results based on the bivariate model for CR Prep can be found in the supplementary file.
\\
\indent A nice side effect of using random effect censored regression methods, as a way to estimate the average cluster expression while adjusting for the LOD, is that it offers the possibility to use the estimate of the random effect variance as a criterion to screen out highly predictive proteins.
We presented two different approaches for performing variable selection based on the estimate of the random effect variance. In the first approach, the decision on which variables to retain is based solely on the magnitude of the random intercept variance $\tau^2$ and is independent of the class outcome. On the other hand, one can choose to determine the optimal fraction of variables to be retained by optimizing the loss function through the use of cross-validation, as demonstrated in section 4.4.2. Variable selection methods based on optimizing a chosen risk function by looking at the class outcome have seen many applications and publications, with Lasso regularization being among the most popular ones. A formal comparison between the various variable selection methods and the here proposed approach falls beyond the scope of this work.
\\
\indent Apart from \textit{a priori} variable selection based either on a fixed fraction of clusters with minimal $\tau^2$ or on selecting the optimal fraction of clusters to be retained through the use of cross validation, one could think of alternative ways to determine a reasonable, fixed across clusters, value for $\tau^2$ in order to get closer to the idea of prediction. One option towards that direction would be to treat $\tau^2$ as a tuning parameter to be optimized via cross-validation. In this way, the amount of shrinkage of the intensity levels is estimated directly from a predictive point of view. We leave the idea of determining the optimal value of the random effect variance via optimization techniques, as an interesting topic of future research.


\section{Conclusion}
We have demonstrated that censored regression can be used successfully to handle the LOD problem in determining the average intensity of isotope clusters in mass-spectrometry proteomic data. In particular in combination with random effects methodology it can contribute to a more efficient preprocessing.

\section*{Acknowledgements}
This work was supported by funding from the European Community's Seventh Framework Programme FP7/2011: Marie Curie Initial Training Network MEDIASRES under the Grant Agreement Number 290025 and by funding from the European Unions Seventh Framework Programme FP7/Health/F5/2012: MIMOmics under the Grant Agreement Number 305280.

\bibliographystyle{SageV}
\bibliography{LOD_arXiv}

\newpage
\section*{Tables}

\bigskip 
\bigskip 
\bigskip 
\bigskip 
\bigskip 
\bigskip 
\bigskip 
\bigskip 
\bigskip 
\bigskip 
\bigskip 
\bigskip 

\begin{table}[h]
\begin{center}
\begin{tabular}{llllllll}
\toprule
&\multicolumn{7}{c}{Validated classification results (based on internal validation)}  \\
\cmidrule(r){2-8}
& CCA & BC & LOD & CR Prep & CR Pred & CR Reest  & TR  \\
\midrule
Error-rate &0.135 (0.008) & 0.125 (0.007)& 0.114 (0.005) &0.110 (0.005)  &0.109 (0.006) & 0.114 (0.008) &0.114 (0.005)  \\
Brier score&0.103 (0.004) & 0.100 (0.003)& 0.085 (0.003) &0.086 (0.003)  &0.084 (0.003) & 0.086 (0.005) &0.087 (0.003)  \\
Deviance   &55.70 (1.85)  & 54.59 (2.18) & 46.56 (2.00)  &47.41 (1.89)   &47.23 (2.01)  &48.12 (2.16)   &48.08 (1.67)   \\
AUC        &0.917 (0.006) & 0.917 (0.009)& 0.943 (0.006) &0.940 (0.006)  &0.942 (0.006) & 0.942 (0.006) &0.940 (0.006)  \\
\bottomrule
\end{tabular}
\end{center}
\caption{Validated classification results (and standard errors) based on complete case analysis (CCA), binary coding (BC), LOD imputation (LOD), random censored regression as preprocessing tool (CR Prep), random censored regression as prediction tool (CR Pred), random censored regression re-estimated (CR Reest) and the ``truth'' (TR).}
\end{table}

\clearpage

\begin{table}
\begin{center}
\begin{tabular}{l p{1.25cm} p{1.25cm} p{1.25cm} p{1.25cm} p{1.25cm} p{1.25cm} p{1.25cm}}
\toprule
&\multicolumn{7}{c}{Validated classification results (based on external validation)}  \\
\cmidrule(r){2-8}
& CCA & BC & LOD & CR Prep & CR Pred & CR Reest  & TR  \\
\midrule
Error-rate  &0.135  &0.125  & 0.115  & 0.087  & 0.096  &  0.076  & 0.106  \\
Brier score &0.113  &0.107  & 0.082  & 0.082  & 0.082  &  0.064  & 0.079  \\
Deviance    &78.01  &70.50  & 58.03  & 56.99  & 57.64  &  47.24  & 54.62  \\
AUC         &0.905  &0.939  & 0.956  & 0.970  & 0.967  &  0.971  & 0.970  \\
\bottomrule
\end{tabular}
\end{center}
\caption{Validated classification results based on complete case analysis (CCA), binary coding (BC), LOD imputation (LOD), random censored regression as preprocessing tool (CR Prep), random censored regression as prediction tool (CR Pred), random censored regression re-estimated (CR Reest) and the ``truth'' (TR).}
\end{table}

\clearpage

\begin{table}
\begin{center}
\begin{tabular}{l p{1.75cm} p{1.75cm} p{1.75cm} p{1.25cm} p{1.25cm} p{1.25cm} }
\toprule
&\multicolumn{6}{c}{Validated classification results}  \\
\cmidrule(r){2-7}
&\multicolumn{3}{c}{Internal validation} & \multicolumn{3}{c}{External validation} \\
\cmidrule(r){2-4}
\cmidrule(r){5-7}
& CR Prep & CR Pred & CR Reest  & CR Prep & CR Pred & CR Reest    \\
\midrule
Error-rate  &0.115 (0.007) &0.115 (0.005) & 0.119 (0.010) & 0.115  & 0.086  &  0.086    \\
Brier score &0.089 (0.004) &0.085 (0.003) & 0.094 (0.007) & 0.085  & 0.084  &  0.081    \\
Deviance    &49.89 (2.16) &48.23 (2.24) & 50.19 (4.09) & 60.12  & 59.34  &  56.95    \\
AUC         &0.934 (0.005) &0.939 (0.006) & 0.939 (0.006) & 0.964  & 0.959  &  0.967    \\
\bottomrule
\end{tabular}
\end{center}
\caption{Validated classification results based on random censored regression with no shrinkage as preprocessing tool (CR Prep), random censored regression with no shrinkage as prediction tool (CR Pred), random censored regression with no shrinkage re-estimated (CR Reest) for internal validation (left part) and external validation (right part).}
\end{table}

\clearpage
\newpage
\section*{Figures}

\bigskip
\bigskip
\bigskip
\bigskip
\bigskip
\bigskip
\bigskip
\bigskip 
\begin{figure}[h]
    \centering
      \includegraphics[height=9cm]{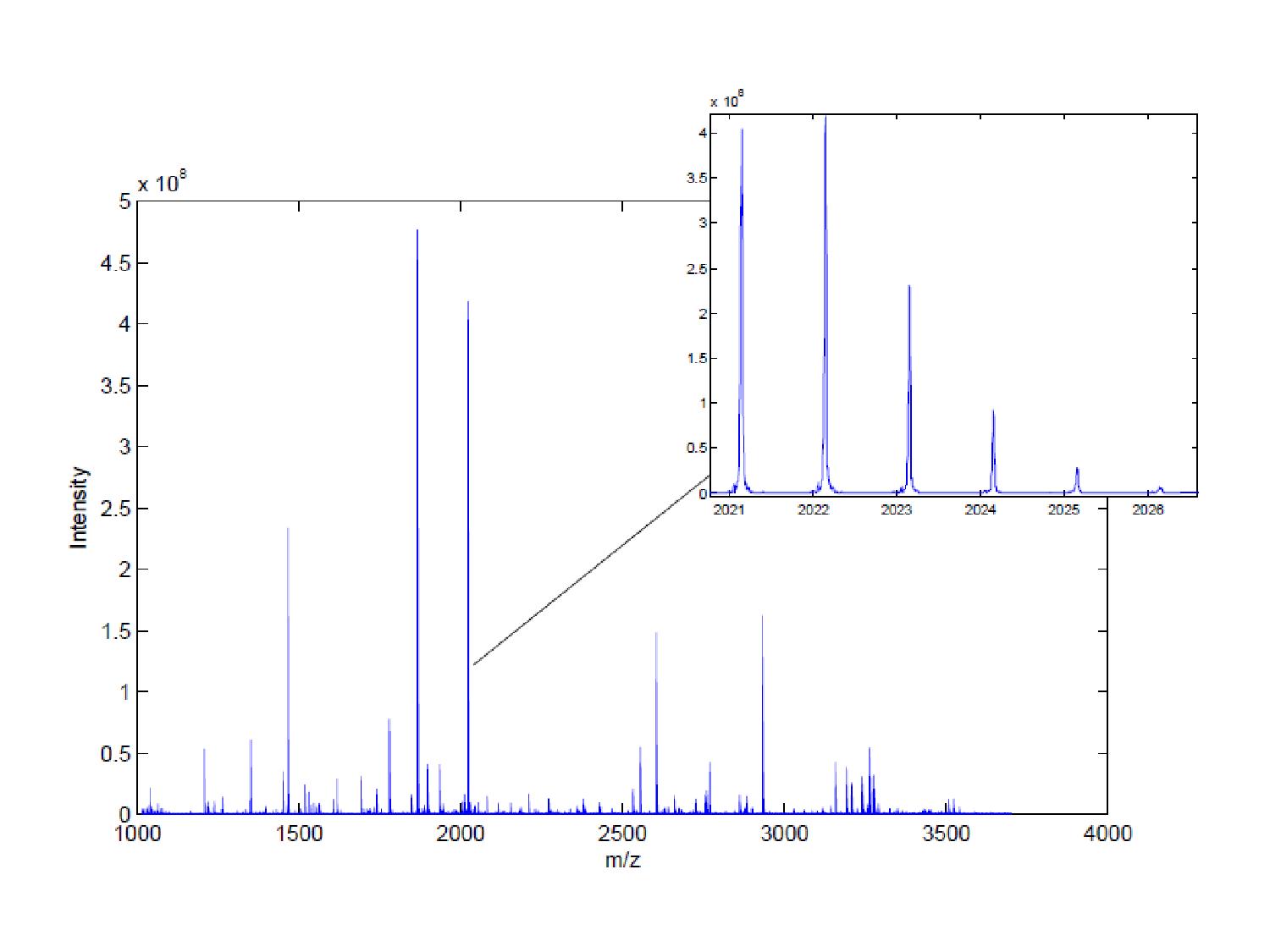}
      \caption{The mass spectrum of a single individual.
      Superimposed is shown an isotopic cluster at position \textit{m/z} 2021,2. }
\end{figure}

\begin{figure}
\centering
\includegraphics[trim=0cm 0.5cm 0cm 1.8cm,height=4.5cm,clip]{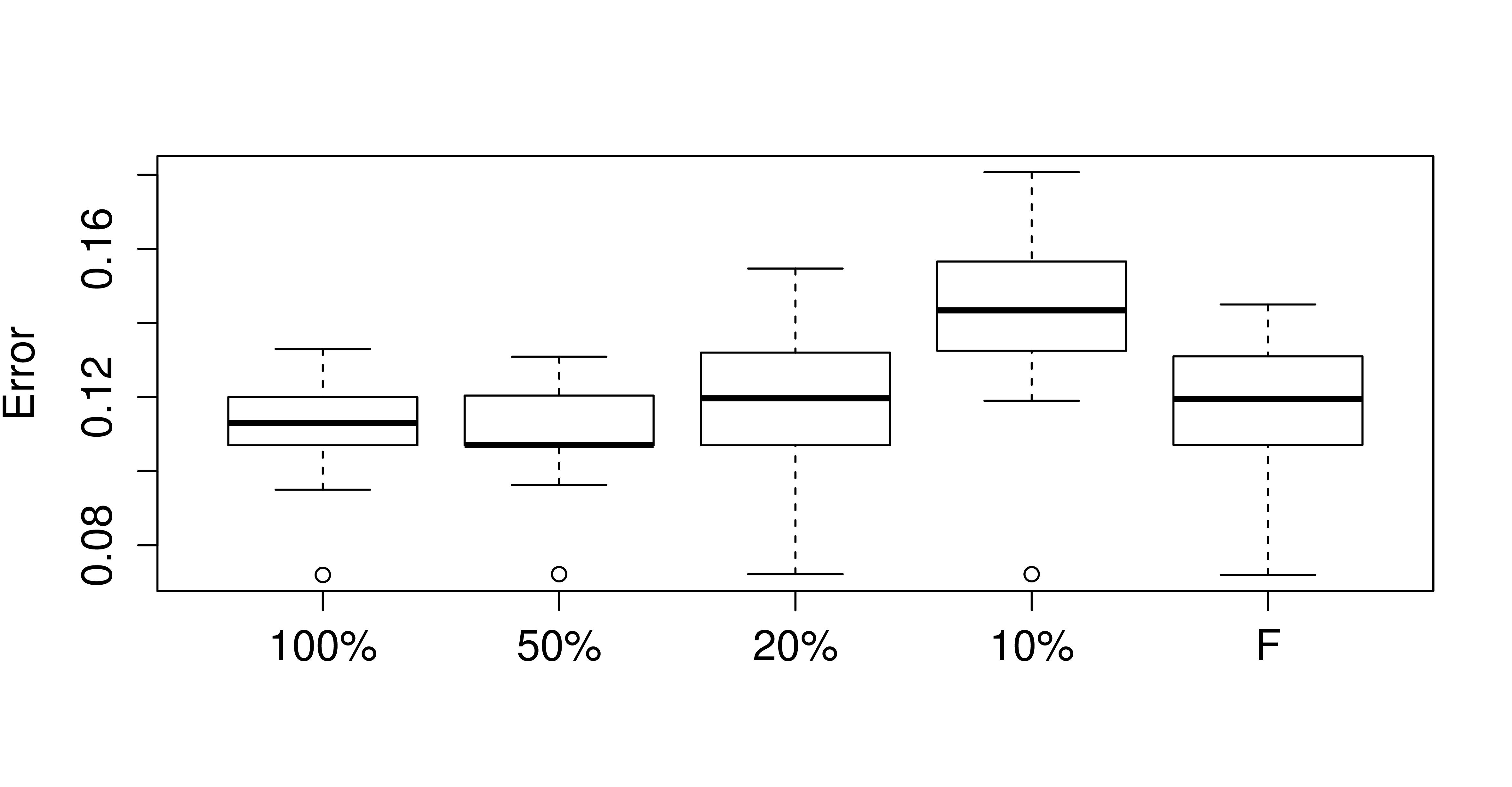}\\
\includegraphics[trim=0cm 0.5cm 0cm 1.8cm,height=4.5cm,clip]{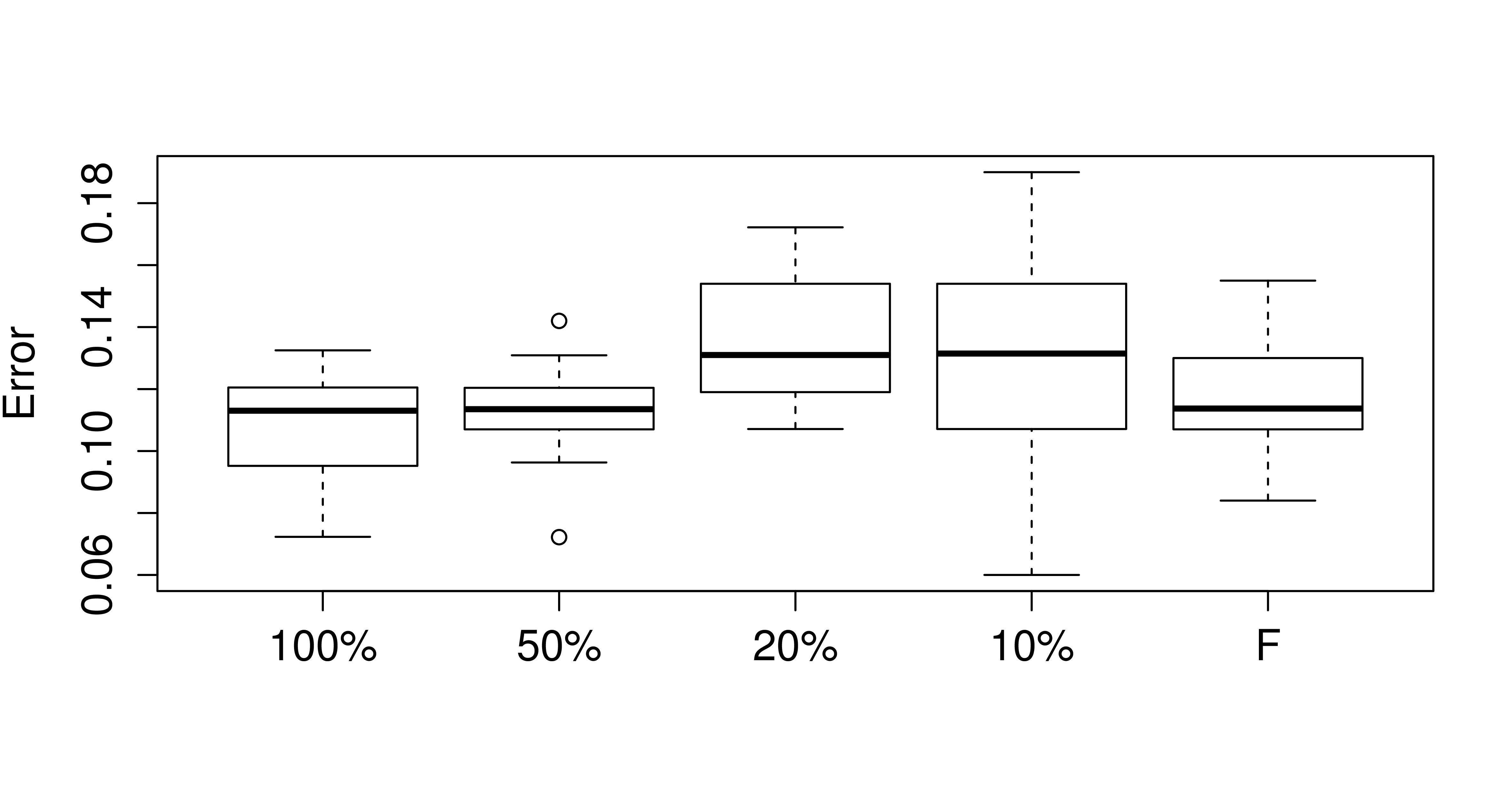}\\
\includegraphics[trim=0cm 0.5cm 0cm 1.8cm,height=4.5cm,clip]{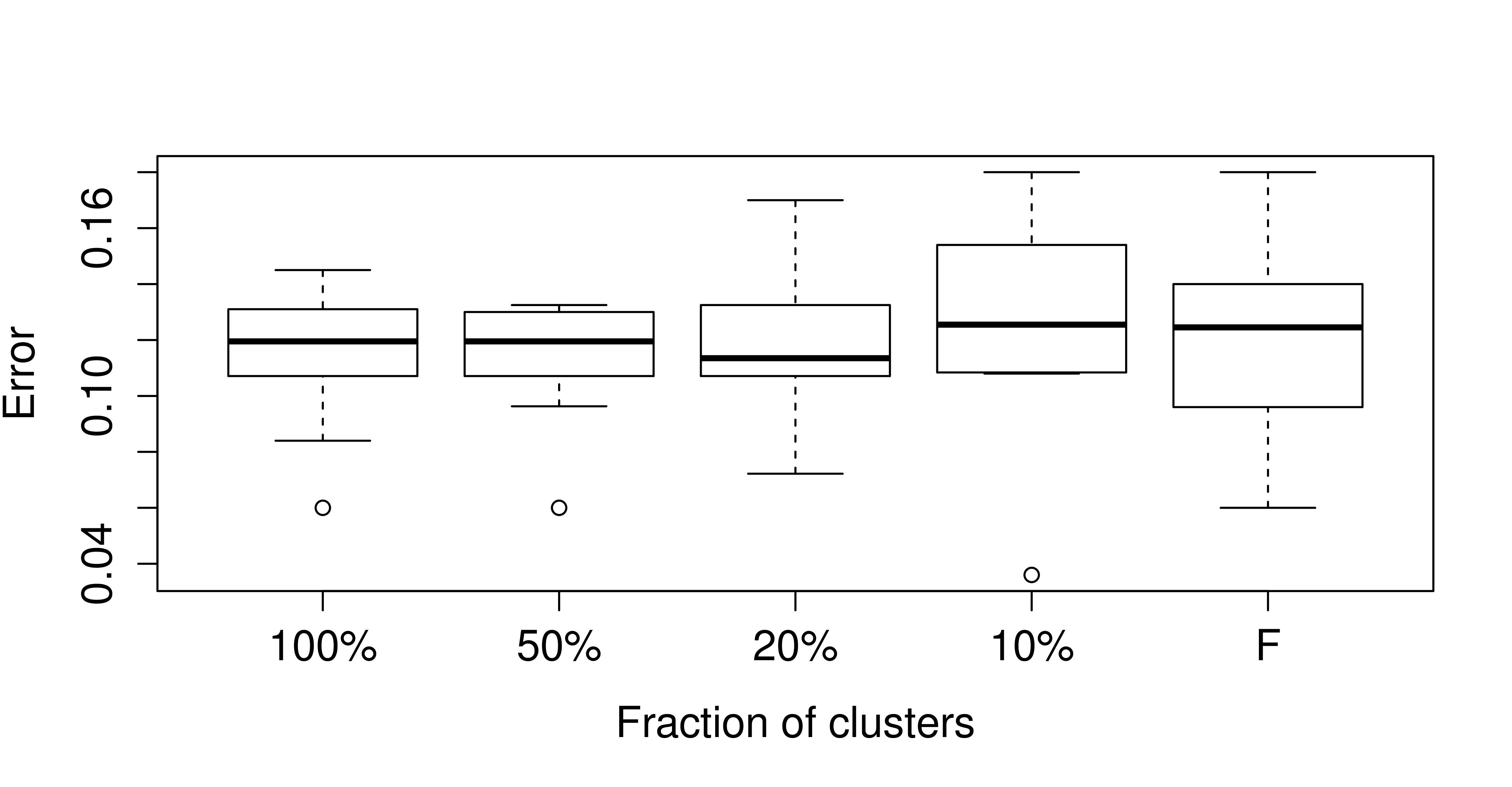}
\caption{Boxplots of error-rates when keeping all clusters, 50\%, 20\%, 10\% and $F$ (=optimal fraction of selected clusters defined based on cross-validation) of total clusters with minimal $\tau_c^2$ based on CR Prep (upper plot), CR Pred (middle plot) and CR Reest (lower plot) for the 10 re-sampled validation sets.}
\end{figure}

\begin{figure}
\centering
\includegraphics[height=6cm]{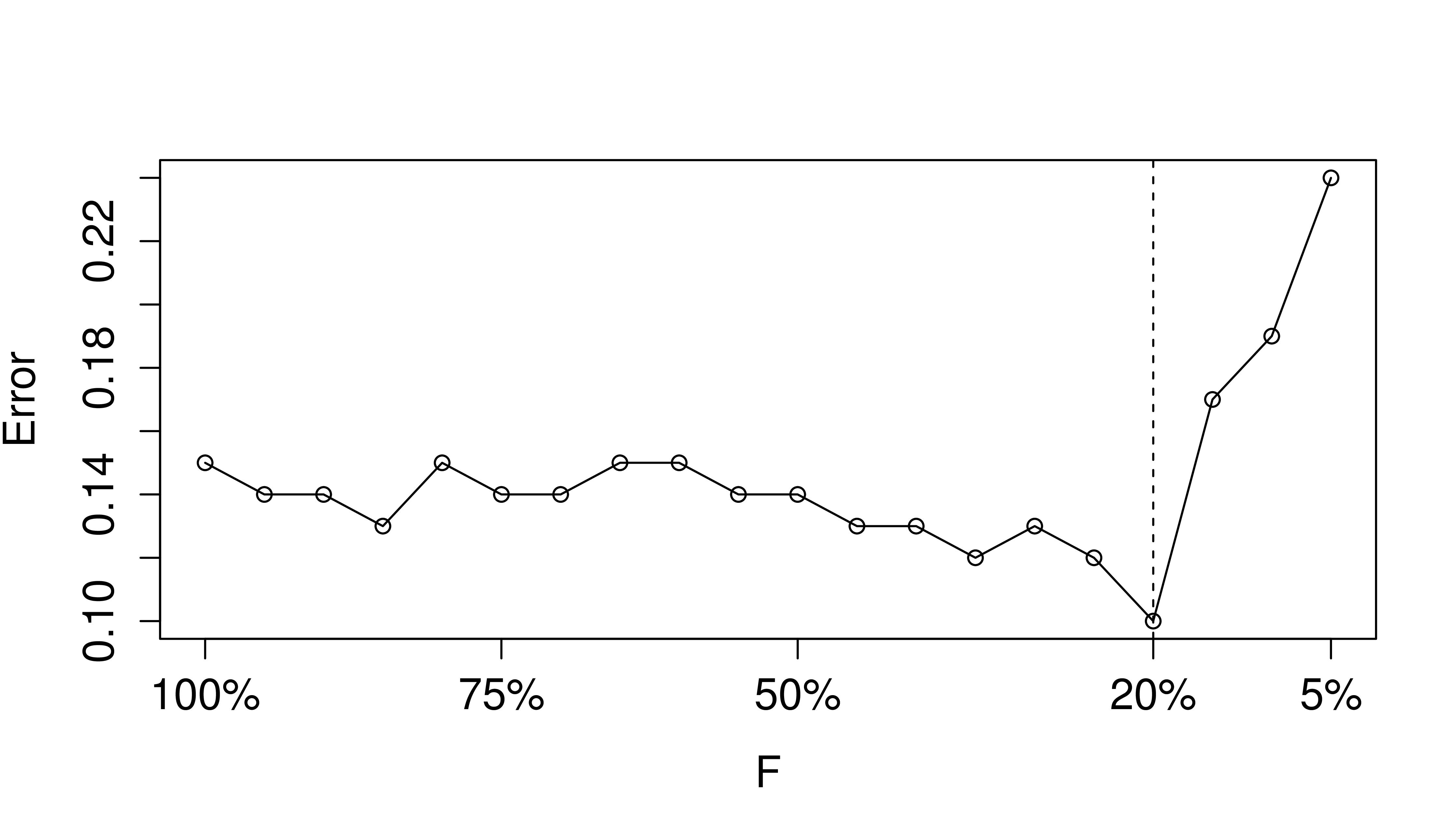}
\caption{Cross-validated error-rate based on CR Prep as the fraction of selected clusters $F$ becomes smaller. Optimal solution is chosen for $F=20\%$ resulting in a subset of just 543 clusters/proteins.}
\end{figure}

\end{document}